\documentclass[12pt]{JHEP3}
\usepackage{amsmath, amscd, amsmath, amsthm, amssymb, xy, xypic}
\usepackage{amssymb}
\usepackage{epsfig}
\usepackage{longtable}

\newtheorem{theorem}{Theorem}[section]

\newtheorem{cor}[theorem]{Corollary}

\newtheorem{example}[theorem]{Example}

\numberwithin{equation}{section}

\def \hd #1 {\bfseries #1  \mdseries}
\def \italic #1 {\bfseries \it #1 \rm \mdseries}
\def \ra {\rightarrow}
\def \cen #1 { \begin{center} #1 \end{center}}

\def \mbz {\mathbb Z}

\def \mbc {\mathbb C}

\def \mco  {\mathcal {O}}

\def \SU {{\rm{SU}}}
\def \setminus {-}
\def \CP {\mathbb{C}\mathbb{P}}

\ifpdf
\else
\usepackage{colordvi}
\fi

\numberwithin{equation}{section}

\title{Calabi--Yau manifolds from pairs of non-compact Calabi--Yau manifolds}

\author{
Nam-Hoon Lee
\\
Department of Mathematics Education, Hongik University,
42-1, Sangsu-Dong, Mapo-Gu,
Seoul 121-791, South Korea
\\
E-mail: \email{nhlee@hongik.ac.kr},
\\
School of Mathematics, Korea Institute for Advanced Study,
\\
Dongdaemun-gu, Seoul 130-722, South Korea
\\
E-mail: \email{nhlee@kias.re.kr}
}

\abstract{
Most of Calabi--Yau manifolds that have been considered by
physicists are complete intersection Calabi--Yau manifolds of
toric
  varieties or some quotients of product types.
 Purpose of this paper is to introduce  a different and rather new kind of construction method of Calabi--Yau manifolds by pasting two \emph{non-compact} Calabi--Yau manifolds.
  We will also in some details explain a curious and mysterious similarity with construction of some $G_2$-manifolds (also called Joyce manifolds), which are  base spaces for
  M-theory.
}


\preprint{}

\begin{document}

\section{Introduction}

There has been great interest in the geometry of Calabi--Yau
manifolds related with string theory.
  Most of physical theories have been investigated for complete intersection Calabi--Yau manifolds of toric varieties or some quotient Calabi--Yau manifolds of products of elliptic curves, $K3$ surfaces or abelian varieties.

 Purpose of this paper is to introduce to physicists a different and rather new kind of construction method of Calabi--Yau manifolds.
  It is by pasting two \emph{non-compact} Calabi--Yau manifolds.
  Those two non-compact Calabi--Yau manifolds are obtained from Fano manifolds, for which the projective spaces are the most typical examples.

  By pasting two non-compact Calabi--Yau threefolds obtained from two $\CP^3$'s, we construct some examples of Calabi--Yau threefolds.   The Hodge numbers of those Calabi--Yau threefolds are given in Table 1.
  There has been found curious and mysterious similarity with construction of some $G_2$-manifolds, which are  base spaces for M-theory of $N=1$ supersymmetry. We will discuss this points also in some details. Finally we remark that the quantum cohomomology and Gromov--Witten invariant calculation method also has been developed(\cite{MuPa}).

 The procedure of  pasting can be also understood as `smoothing normal crossing varieties', which are  subjects of algebraic geometry.
  The conditions for pasting can be stated rather more conveniently and rigorously in algebro-geometric terminologies.
 So we will first describe the method of smoothing normal crossing varieties, which is mathematically equivalent to the topological pasting.
 Later we will explain the topological pasting procedure (Section \ref{top}).

\section{Some terminologies}

In this paper, a \italic{Calabi--Yau $n$-fold} means a compact
K\"ahler manifold of dimension $n$ with trivial canonical class
such that the Hodge numbers are: $h^{i,0} =0$ for $0 < i <n$. The
triviality of the canonical class can be replaced by the condition
that it admits  Riemannian metric with $\SU(n)$ as its holonomy
group. In this definition, a $K3$ surface is a Calabi--Yau
twofold. A $G_2$-manifold (also called Joyce manifold) is a
Riemannian manifold of real 7-dimension whose holonomy group is
$G_2$ group (see \cite{Jo}).
 A Fano manifold is a smooth projective variety with
ample anticanonical class.  The most typical examples are
$\CP^n$'s. Fano manifolds are well-understood and completely
classified for dimensions less than four (see \cite{IsPr} for
example). There are 105 families of Fano threefolds.

A normal crossing varieties is a union of irreducible varieties,
intersecting with another in a specific way.  We will only
consider following type of normal crossing varieties:
$$X_0 = Y_1 \cup Y_2,$$
where $Y_1, Y_2$ are smooth varieties of dimension $n$ and $D:=Y_1
\cap Y_2$ is a smooth variety of dimension $n-1$. We require that
$X_0$ has singularities along $D$, locally isomorphic to
\begin{align} \label{sm}
\{x=(x_1, x_2, \cdots, x_n) \in \mbc^{n+1} \big{|} x_1 x_2 = 0 \}.
\end{align}

The set $\{x\big|x_i=0\}$ corresponds to $Y_i$ for $i=1, 2$ and
the set
$$\{x\big |x_1=0, x_2=0\}$$
 corresponds to $D$.
The varieties $Y_1, Y_2$ will be Fano manifolds or some blow-ups
of them and $D$ will be a Calabi--Yau $(n-1)$-fold.

Now let us explain what smoothing is. Let $\pi: X \ra  \Delta$ be
a proper map from a  K\"ahler $(n+1)$-fold $X$ onto the unit disk
$$ \Delta = \{ t \in \mbc \big| \|t\| \leq 1 \}$$
 such that the fibers $X_t = \pi^{-1} (t)$ are  manifolds of dimension $n$ for every $t \neq 0$ (generic) and the central
fiber
$$X_0=\pi^{-1} (0) = Y_1 \cup Y_2$$
 is a normal
crossing variety. We denote the generic fiber by $X_t$. The
condition, $t \neq 0$, is assumed in this notation. We call such a
map $\pi$ (or simply the total space $X$) a semi-stable
degeneration of $X_t$ and  $X_0$ the central fiber. In practice,
we say that the smooth manifold $X_t$ degenerates to the normal
crossing variety $X_0$ as $t$ goes to zero. Conversely $X_t$ is
said to be a \emph{smoothing} of $X_0$ or $X_0$ is smoothable to
$X_t$. For different non-zero values $t$ and  $t'$, $X_t$ and
$X_{t'}$ are diffeomorphic to each other.

The process of resolution of singularities of $X_0$ by smoothing
is as follows: In the equation (\ref{sm}), it has singularities,
locally isomorphic to
$$\{(x_1, x_2, \cdots, x_n) \in \mbc^n \big | x_1 x_2 = t \}$$
 with $t=0$, which corresponds to degenerated fiber. Smoothing is letting $t$ go apart from $0$. Then the singularities instantly disappear, which corresponds to the generic fiber $X_t$.

\section{The smoothing conditions}
The general reference for this section is \cite{Lee}. Given a
normal crossing variety, its smoothing is not always possible. We
need some conditions. Those conditions are stated in the following
smoothing theorem, which is a corollary of a theorem of Y.
Kawamata and Y. Namikawa (\cite{KaNa}).

\begin{theorem} \label{kana}Let $X_0=Y_1 \cup Y_2$ be a normal crossing two projective manifolds $Y_1, Y_2$ of dimension $n \geq 3$ such that:
\begin{enumerate}
\item Let $D = Y_1\cap Y_2$, then $D \in |{-}K_{Y_i}|$ for
    $i=1$ and $2$. So $D$ is an anticanonical section of $Y_1$
    and $Y_2$
\item The Hodge numbers  are $h^{j, 0} ({Y_i}) = h^{j, 0}
    (D)=0$ for $0< j < n-1$. Note that this condition,
    together with (1),  implies that $D$ is a Calabi--Yau
    $(n-1)$-fold.
\item There are ample divisors $H_1$, $H_2$ on $Y_1$, $Y_2$
    respectively such that $H_1|_D$ is linearly equivalent to
    $H_2|_D$.
\item The divisor $-K_{Y_1}|_D  -K_{Y_1}|_D$ on $D$ is
    linearly equivalent to the zero divisor on $D$.   This
    condition is called d-semistability.
\end{enumerate}
Then $X_0$ is smoothable to an  $n$-fold $X_t$ with trivial
canonical class $K_{X_t} = 0$.
 \end{theorem}
One can show that  $h^{i,0}({X_t})=0$ for $0<i<n$. Accordingly
$X_t$ is a Calabi--Yau $n$-fold.

If $Y_1$ and $Y_2$ are Fano manifolds, they satisfies the
condition (2) in the above theorem and very often they have a
Calabi--Yau $(n-1)$-fold as their anticanonical sections, which is
related with the condition (1). So now take a Fano $n$-fold $Z$
with a smooth anticanonical section $D$. Let $Z_1$ and $Z_2$ are
copies of $Z$. Let $Z_0 = Z_1 \cup_D Z_2$, where `$\cup_D$' means
pasting along $D$ (Note that $Z_1$ and $Z_2$ contain  copies of
$D$). Now $Z_0$ is a normal crossing and one can check that it
satisfies the conditions (1), (2) and (3). However the divisor
$$ -K_{Y_1}|_D  -K_{Y_1}|_D$$
is \emph{not} linearly equivalent to the zero divisor on $D$. So
$Z_0$ does not satisfy the condition (4). Accordingly the above
theorem is not applicable to $Z_0$. To remedy the situation, we
blow up one of $Z_1, Z_2$ along some subvarieties on $D$. Let us
choose a smooth irreducible divisor $c$ (the general
non-irreducible  case will be discussed later) from the complete
linear system $\big |-K_{Z_1}|_D  -K_{Z_2}|_D \big |$.

Let $Y_1 = Z_1$ and $Y_2$ be the blow-up of $Z_2$ along $c$. Since
$c$ lies on $D$, the blow-up does not change $D$. So $Y_2$ still
contains a copy of $D$ and one can easily verify that $D$ is again
an anticanonical section. Now take $X_0 = Y_1 \cup_D Y_2$, then it
satisfies all conditions in the theorem and therefore smoothable
to a Calabi--Yau $n$-fold $X_t$.

Let us take a simple example.
\begin{example} \label{ex}
 Let $Y_1 = \CP^3$, $D$ be
a smooth quartic hypersuface in $\CP^3$ and $Y_2 $ be the blow-up
of $\CP^3$ along a smooth curve $c \in |\mco_D(8)|$.  Let $X_0 =
Y_1 \cup_D Y_2$, then it is smoothable to a Calabi--Yau threefold
$X_t$. Its Hodge numbers and topological Euler number are (see
Corollary \ref{main})
\begin{align*}
&h^{1, 1} =1,  \\
&h^{1, 2} = h^{2,1}= 149,\\
&e=2(h^{1, 1} - h^{1, 2})=-296.
\end{align*}
\end{example}

Note that the Calabi--Yau manifold $X_t$ is determined by the
quadruple $(Y_1, Y_2, D, c)$. So one can simply say that for a
quadruple $(Y_1, Y_2, D, c)$, there is a Calabi--Yau manifold
obtained by smoothing.

There are two possibilities of generalization. Firstly $Z_1$ and
$Z_2$ do not need to be the same Fano threefold. So let $Z_1$ and
$Z_2$ be Fano $n$-folds which contains a common smooth
anticanonical section $D$ in their anticanonical section. Secondly
the divisor $c$ does not need to be irreducible. Let
$$c=c_1 + c_2+ \cdots + c_r$$
 be a divisor that is composed of smooth divisors $c_i$'s.
We remark that these $c_i$'s do not need to be distinct. Let
$Y_1=Z_1$ and get $Y_2$ by blowing up successively along with the
\emph{proper transforms} of  $c_1, \cdots, c_r$ in this order.
Note that $Y_2$ still contains a copy of $D$ as its anticanonical
section because all the blow-up centers lies on $D$. We make a
normal crossing variety $X_0 = Y_1 \cup_D Y_2$. Again one can
smooth $X_0$ to a Calabi--Yau $n$-fold $X_t$ if $c$ belongs to
$\big |-K_{Z_1}|_D  -K_{Z_2}|_D \big |$. In summary (\cite{Lee},
Corollary 7.2):

\begin{cor} \label{fano} Let $Z_1$ and $Z_2$ be Fano $n$-folds which contains a common smooth anticanonical
section $D$ in their anticanonical section and $c=c_1 + c_2+
\cdots + c_r$ belong to the linear system  $\big |-K_{Z_1}|_D
-K_{Z_2}|_D \big |$, where $c_i$'s are smooth. Then for such a
quadruple $(Y_1, Y_2, D, c)$, there is a Calabi--Yau $n$-fold
obtained by smoothing and in dimension three, the Hodge numbers
are:
\begin{enumerate}
\item $h^{1, 1} = h^{2} (Y_1) + h^{2} (Y_2) - k -1$, where
$$k = {\rm{rk}} ({\rm{im}} (H^2 (Y_1, \mbz) \oplus H^2 (Y_2, \mbz) \ra H^2 (D, \mbz))).$$
The map is induced by the inclusion $D \hookrightarrow Y_i$.
\item $h^{1, 2} =h^{2,1} = 21 + h^{1, 2} (Y_1) + h^{1, 2}
    (Y_2) - k$.
\end{enumerate}
\end{cor}
Let us remark that the order of blow-ups along $c_i$'s matters
although the Hodge numbers do not depend on it. The cup-product on
$H^2(X_t, \mbz)$ generally depends on the order. So blowing-up in
a different order generally gives Calabi--Yau manifolds of
different topological types (\cite{Lee}, Section 7).

\section{Calabi--Yau manifolds from pairs of $\CP^3$'s}
There are 105 families of Fano threefolds and far more families
for fourfolds (see \cite{IsPr} for example). Paring them and
applying Corollary \ref{fano}, one can construct many examples of
Calabi--Yau manifolds. Projective spaces are the most elementary
examples of Fano manifolds. In this section, we will compute Hodge
numbers of Calabi--Yau manifolds obtained from pairs of $\CP^3$'s.

Let $D$ be any smooth hypersurface of degree $n+1$ in $\CP^n$,
then $D$ is an anticanonical section of $\CP^n$. As before choose
divisor $c=c_1 + \cdots + c_r$ of $D$. Let $Y_1=\CP^n$ and get
$Y_2$ by blowing up $\CP^n$. If $c$ is of degree $2n$, $X_0 = Y_1
\cup_D Y_2$ is smoothable to a Calabi--Yau $n$-fold $X_t$. From
Corollary \ref{kana},  one can say:

\begin{cor}\label{proj}Let $D$ be any smooth hypersurface of degree $n+1$ in $\CP^n$ and $c=c_1 + \cdots + c_r$ be a divisor of degree $2n+2$ on $D$, where $c_i$'s are smooth. Then for such triple $(\CP^n, D, c)$ there is a corresponding Calabi--Yau $n$-fold obtained by smoothing.
\end{cor}

Let us go over the examples of Calabi--Yau threefolds which are
constructed from two copies of $\CP^3$ (\cite{KaNa}). Let $D$ be a
smooth quartic $K3$ surface in $\CP^3$, defined by
$$D = \{x_0^4 + x_1^4 + x_2^4 + x_3^4  = 0\} \subset \CP^3.$$
It is the Fermat quartic $K3$ surface and we can find various
divisors on it.
\begin{enumerate}
\item For a primitive 8-th root of unity $\xi$, define the
    divisor $\Gamma_{i, j, k}$ of $D$ as \cen{$\Gamma_{i, j,
    k}  =  \{  (x_0, \dots, x_3) \in D \big | x_i = \xi^k
    x_j\} $} Let $L = \{ \Gamma_{0, 2, 1}, \Gamma_{0, 2, 2},
    \Gamma_{0, 2, 3}, \Gamma_{1, 2, 4} \}$. It can be easily
    shown that each of these $\Gamma_{i,j,k}$'s in $L$
    consists of 4 lines which meet at a single point.

\item For the sixteen lines on $D$, coming from those
    $\Gamma_{i,j,k}$'s in $L$, we can assign a divisor $F \in
    |\mco_D (1)|$ as \cen{$ F = D \cap H $,} where $H$ is a
    generic hyperplane which contains the line. Then $F$ is
    composed of the line and a smooth cubic curve.   Let $N$
    be the set of such divisors.
\end{enumerate}

Now we take
$$c = E_1 + \dots + E_s + \Gamma_1 + \dots + \Gamma_a + F_1+
\dots + F_u,$$

 where $E_i$ for $1 \leq i
\leq s$ is degree of $e_i$, $\Gamma_i$ is a member of $L$ and
$F_i$ is a member of  $N$   respectively such that
$$a  +\sum_i e_i  + u = 8.$$
 Then $X_0$ is smoothable and the Hodge numbers
are given by the following corollary, which is a simple
application of Corollary \ref{fano}.

\begin{cor} \label{main}Let $X_t$ be the Calabi--Yau threefold corresponding to the triple $(\CP^3, D, c)$
 then the Hodge numbers of $X_t$ are:\
\begin{align*}
h^{1, 1} &=  s + 4 a + 2 u,  \\
h^{1, 2} &=h^{2,1}= 20 + 2\sum_i e_i^2 + s + 4 a + 2 u.
\end{align*}

\end{cor}

Table 1 in the appendix is the exhaustive list of Hodge numbers of
Calabi--Yau threefolds constructible in this way. For example, if
one take $a=b=0$, $s=1, t=0, e_1=8$, and $u=v=0$,
\begin{align*}
&h^{1, 1} =  1 + 0 + 0=1,  \\
&h^{1, 2} = h^{2,1} = 20 + 2 \cdot 64 + 1+0 + 0= 149,\\
&e=2(h^{1, 1} - h^{1, 2})=-296,
\end{align*}
where $e$ is the topological Euler number. This is Example
\ref{ex}. Note that there are many examples with the same Euler
number that have different Hodge numbers in the table.

 \section{Topological interpretation and  $G_2$-manifolds} \label{top}
 The general references for this section are \cite{Ty} and \cite{Ko}.
As stated in the introduction, one can obtain $X_t$ by pasting
two non-compact Calabi--Yau manifolds. We mean a non-compact
Calabi--Yau manifold by a quasi-projective variety $Y^*$ such that
$$Y^* = Y - D,$$
where $Y$ is a projective manifold and $D$ is a smooth section in
$\big| - K_Y \big|$ (\cite{TiYa1}, \cite{TiYa2}).
 The condition (4) in Theorem \ref{kana} is equivalent to the triviality of the following line bundle on $D$:
$$ N_{D/Y_1}\otimes N_{D/Y_2},$$
where $N_{D/Y_i}$ is the normal bundle to $D$ in $Y_i$. One can
regard $N_{D/Y_i}$ as a complex manifold containing $D$. Then
$$N_{D/Y_i}^* :=  N_{D/Y_i} \setminus D$$
 is a $\mbc^*$-bundle over $D$, where $\mbc^* :=\mbc - \{0\}$. The triviality condition implies the map
 $$\varphi: N_{D/Y_1}^* \ra N_{D/Y_2}^*,$$
locally defined by
 $$(x \in \mbc^* , y \in D) \mapsto \left({1}/{x} , y \right),$$
 is globally well-defined and an isomorphism. Note that $D$ in $Y_i$ has a neighborhood $U_i$ that is homeomorphic to $N_{D/Y_i}$.
 Let $U_i^* = U_i - D$. Then the map $\varphi$ induces a homeomorphism between $U_1^*$ and $U_2^*$.
 One can construct a manifold $W$ by pasting together $Y_1^*$ and $Y_2^*$ along $U_1^*$ and $U_2^*$ with the homeomorphism. Here $Y_i^* = Y_i - D$ is a non-compact Calabi--Yau manifolds since $D$ is a anticanonical section of $Y_i$.
 The manifold $W$ is homeomorphic to the Calabi--Yau manifold $X_t$ obtained by smoothing $X_0$.
 So the Calabi--Yau manifold $X_t$ can be obtained by topologically pasting non-compact Calabi--Yau manifolds $Y_1^*$ and $Y_2^*$. In summary (\cite{Ty}, Section 3.2),
 \begin{theorem}
 Let $Y_1, Y_2, D$ be as in Theorem \ref{kana}. Then $W$, obtained by  pasting non-compact Calabi--Yau manifolds $Y_1^*$ and $Y_2^*$, homeomorphic
 to the  manifold $X_t$.
 \end{theorem}

A construction method of $G_2$-manifolds has been proposed by A.
Kovalev (\cite{Ko}). It is a technique of so-called `twisted
connected sum'. It also starts with two non-compact Calabi--Yau
threefolds. The condition here is that the normal normal bundles $
N_{D/Y_1}$, $N_{D/Y_2}$ are all trivial, which is stronger that
the condition (4) in Theorem \ref{kana}. In identifying $D$'s in
$Y_1$ and $Y_2$, it uses hyperk\"ahler isometries. By the
triviality of the normal bundles, $U_i^*$ is homeomorphic to
$\mbc^* \times D$. Consider $\hat Y_i = Y_i -U_i$. Then $\hat Y_i$
is a manifolds with boundary $B_i$ that is homeomorphic to $S^1
\times D$. As before we make a homeomorphism from $B_1 \times S^1$
to $B_2 \times S^1$ by
$$(x\in S^1, y\in D, z\in S^1) \mapsto (z, y, x).$$
Note that the variables $x$, $z$ are interchanged. This is why the
word `twisted' comes up. Now make a manifold $M$ by pasting
together $\hat Y_1 \times S^1$ and $\hat Y_2 \times S^1 $ along
$B_1 \times S^1$ and $B_2 \times S^1$ with the homeomorphism. Then
$M$ is a real 7-dimensional compact manifold. A. Kovalev showed
(\cite{Ko}):
 \begin{theorem}
 The manifold $M$ admits a Riemannian metric with $G_2$ holonomy group.
 \end{theorem}
 In this method, one can construct hundreds examples of $G_2$ manifolds (\cite{Ko}, \cite{KoLe}).

Roughly speaking, two non-compact Calabi--Yau threefolds give rise
to both of a Calabi--Yau threefold and a $G_2$-manifold, which are
mathematically different object. However it seems that there
should be a close connection in the physical theories realized on
them. It seems  a natural and interesting question how the
superstring theory on the Calabi--Yau threefold and M-theory on
the $G_2$-manifold are related.

An anonymous  referee pointed out that this construction looks closely related to the subject
of Type II -- heterotic duality in string theory. According to the referee, the `tops' and `bottoms' that
appear in \cite{CaSk} and \cite{CaFo}, for example, seem closely
linked to the construction, at least in simple cases.


\appendix

\begin{table}[h]
$$
\begin{array}{|rrr||rrr||rrr||rrr||rrr|} \cline{1-3}
\, & \,& &\\
h^{1, 1}&  h^{2, 1}&   e & \\ \hline

\tiny
  1&    149&    -296 & 2&   122&    -240 & 2&   102&    -200 & 2&   90& -176 & 2&  86& -168 \\ \hline

3&  120&    -234 & 3&   99& -192 & 3&   83& -160 & 3&   75& -144 &
3&   71& -136 \\ \hline

3&  67& -128 & 4&   80& -152 & 4&   73& -138 & 4&   68& -128 & 4&
64& -120 \\ \hline

4&  60& -112 & 4&   56& -104 & 5&   119& -228 & 5&   95& -180 & 5&
66& -122 \\ \hline

5&  65& -120 & 5&   58& -106 & 5&   57& -104 & 5&   53& -96 & 6&
76& -140 \\ \hline

6&  72& -132 & 6&   64& -116 & 6&   63& -114 & 6&   60& -108 & 6&
55& -98 \\ \hline

6&  54& -96 & 6&    51& -90 & 6&    50& -88 & 6&    36& -60 & 7&
94& -174 \\ \hline

7&  74& -134 & 7&   65& -116 & 7&   61& -108 & 7&   61& -108 & 7&
57& -100 \\ \hline

7&  53& -92 & 7&    52& -90 & 7&    49& -84 & 7&    48& -82 & 7&
47& -80 \\ \hline

8&  75& -134 & 8&   63& -110 & 8&   62& -108 & 8&   59& -102 & 8&
59& -102 \\ \hline

8&  54& -92 & 8&    51& -86 & 8&    50& -84 & 8&    50& -84 & 8&
46& -76 \\ \hline

8&  45& -74 & 8&    44& -72 & 9&    93& -168 & 9&   73& -128 & 9&
60& -102 \\ \hline

9&  57& -96 & 9&    52& -86 & 9&    51& -84 & 9&    48& -78 & 9&
48& -78 \\ \hline

9&  47& -76 & 9&    44& -70 & 9&    43& -68 & 9&    42& -66 & 10&
74& -128 \\ \hline

10& 62& -104 & 10&  58& -96 & 10&   50& -80 & 10&   49& -78 & 10&
46& -72 \\ \hline

10& 45& -70 & 10&   44& -68 & 10&   42& -64 & 10&   41& -62 & 10&
40& -60 \\ \hline

11& 72& -122 & 11&  59& -96 & 11&   59& -96 & 11&   56& -90 & 11&
51& -80 \\ \hline

11& 47& -72 & 11&   47& -72 & 11&   44& -66 & 11&   43& -64 & 11&
42& -62 \\ \hline

11& 39& -56 & 11&   38& -54 & 12&   57& -90 & 12&   49& -74 & 12&
48& -72 \\ \hline

12& 45& -66 & 12&   44& -64 & 12&   41& -58 & 12&   40& -56 & 12&
39& -54 \\ \hline

12& 37& -50 & 12&   36& -48 & 13&   71& -116 & 13&  55& -84 & 13&
46& -66 \\ \hline

13& 43& -60 & 13&   42& -58 & 13&   41& -56 & 13&   38& -50 & 13&
37& -48 \\ \hline

13& 35& -44 & 13&   34& -42 & 14&   56& -84 & 14&   48& -68 & 14&
44& -60 \\ \hline

14& 40& -52 & 14&   39& -50 & 14&   38& -48 & 14&   36& -44 & 14&
35& -42 \\ \hline

14& 32& -36 & 15&   54& -78 & 15&   45& -60 & 15&   42& -54 & 15&
41& -52 \\ \hline

15& 37& -44 & 15&   36& -42 & 15&   34& -38 & 15&   33& -36 & 15&
30& -30 \\ \hline

16& 43& -54 & 16&   39& -46 & 16&   38& -44 & 16&   35& -38 & 16&
34& -36 \\ \hline

16& 31& -30 & 16&   28& -24 & 17&   53& -72 & 17&   41& -48 & 17&
36& -38 \\ \hline

17& 35& -36 & 17&   33& -32 & 17&   32& -30 & 17&   29& -24 & 18&
42& -48 \\ \hline

18& 38& -40 & 18&   34& -32 & 18&   33& -30 & 18&   30& -24 & 18&
27& -18 \\ \hline

19& 40& -42 & 19&   35& -32 & 19&   32& -26 & 19&   31& -24 & 19&
28& -18 \\ \hline

20& 33& -26 & 20&   32& -24 & 20&   29& -18 & 20&   26& -12 & 21&
39& -36 \\ \hline

21& 31& -20 & 21&   30& -18 & 21&   27& -12 & 22&   32& -20 & 22&
28& -12 \\ \hline

22& 25& -6 & 23&    30& -14 & 23&   29& -12 & 23&   26& -6 & 24&
27& -6 \\ \hline

24& 24& 0 & 25& 29& -8 & 25&    25& 0 & 26& 23& 6 & 27& 24& 6 \\
\hline

28& 28& 0 & 28& 22& 12 & 29&    23& 12 & 30&    21& 18 & 32& 20&
24 \\ \hline
\end{array}
$$
\label{table}
\caption{Hodge numbers of Calabi--Yau threefolds}
\end{table}

\normalsize

\end{document}